\documentclass[aps,pra,twocolumn]{revtex4}
\usepackage{amssymb}
\usepackage{amsfonts}
\usepackage{amsmath}
\usepackage{graphicx}
\usepackage{bbm}

\begin{document}

\title{Detection of Bidirectional System-Environment Information Exchanges}
\author{Adri\'an A. Budini}
\affiliation{Consejo Nacional de Investigaciones Cient\'{\i}ficas y T\'{e}cnicas
(CONICET), Centro At\'{o}mico Bariloche, Avenida E. Bustillo Km 9.5, (8400)
Bariloche, Argentina, and Universidad Tecnol\'{o}gica Nacional (UTN-FRBA),
Fanny Newbery 111, (8400) Bariloche, Argentina}
\date{\today }

\begin{abstract}
Quantum memory effects can be related to a bidirectional exchange of
information between an open system and its environment, which in turn
modifies the state and dynamical behavior of the last one. Nevertheless,
non-Markovianity can also be induced by environments whose dynamics is not
affected during the system evolution, implying the absence of any physical
information exchange. An unsolved open problem in the formulation of quantum
memory measures is the apparent impossibility of discerning between both
paradigmatic cases. Here, we present an operational scheme that, based on
the outcomes of successive measurements processes performed over the system
of interest, allows to distinguishing between both kinds of memory effects.
The method accurately detects bidirectional information flows in diverse
dissipative and dephasing non-Markovian open system dynamics.
\end{abstract}

\maketitle

\section{Introduction}

In its modern conception, quantum non-Markovianity \cite%
{breuerbook,vega,wiseman} is related to a twofold exchange of information
between an open system and its environment \cite{BreuerReview,plenioReview}.
Over the basis of unitary system-environment models, it is commonly assumed
that this bidirectional informational flow (BIF) is mediated by physical
processes that modify the state and dynamical behavior of the environment.
In spite of the consistence of this picture \cite%
{EnergyBackFLow,Energy,HeatBackFlow}, it is well known that memory effects
can also be induced by reservoirs whose state and dynamical behavior are not
affected at all by its coupling with the open system. Evidently, this
feature implies the absence of any \textit{physical} system-bath information
exchange. Stochastic Hamiltonians \cite{cialdi,GaussianNoise,morgado,bordone}%
, incoherent bath fluctuations \cite%
{lindbladrate,boltzman,vasano,PostMarkovian,shabani}, collisional models 
\cite{colisionVacchini,embedding}, and (system) unitary dynamics
characterized by random parameters \cite{ciracR,buchleitner,nori,wudarski}
are some examples of this \textquotedblleft casual
bystander\textquotedblright\ (non-Markovian) environment action. The
environment affects the system dynamics but its (statistical) state is never
influenced by the system.

An open problem in the formulation of quantum non-Markovianity is the lack
of an underlying prescription (based only on system information) able to
discriminate between the previous two complementary cases. In fact, even
when a wide variety of memory witnesses (defined from the system propagator
properties) has been proposed \cite%
{BreuerFirst,cirac,rivas,breuerDecayTLS,fisher,fidelity,dario,mutual,geometrical,DarioSabrina,brasil,sabrina,canonicalCresser,cresser,Acin,indu,poland,chile}
and implemented experimentally \cite%
{BreuerExp,breuerDrift,urrego,khurana,sun,mataloni,pan}, even in absence of
BIFs most of them may inaccurately detect an \textquotedblleft
environment-to-system backflow of information\textquotedblright\ \cite%
{cialdi,GaussianNoise,morgado,bordone,lindbladrate,boltzman,vasano,PostMarkovian,shabani,colisionVacchini,embedding,wudarski,buchleitner}%
. This incongruence emerges because quantum master equations with very
similar structures describe the (non-Markovian) system dynamics in presence
or absence of BIFs.

The previous limitation implies a severe constraint on the classification
and interpretation of memory effects in quantum systems. For example, there
exist non-Markovian dynamics whose underlying memory effects are \textit{%
classified} as \textquotedblleft extreme\textquotedblright\ ones.
Nevertheless, these dynamics emerge from simple classical statistical
mixtures of (memoryless) Markovian system evolutions. Added to the absence
of any physical BIF, the reading of memory effects as quantum ones becomes
meaningless in this situation. Remarkable cases are quantum master equations
with an ever (time-dependent) negative rate (eternal non-Markovianity) \cite%
{canonicalCresser,megier}\ as well as \textquotedblleft maximally
non-Markovian dynamics\textquotedblright\ where the stationary state may
recover the initial condition \cite{DarioSabrina,maximal}. On the other
hand, the \textit{interpretation} of this kind of dynamics in terms of
measurement-based stochastic wave vector evolutions may becomes ambiguous
(Markovian or non-Markovian) by taking into account or not the underling
statistical mixture. In fact, for each Markovian system evolution in the
statistical ensemble one can associate a Markovian stochastic wave vector
evolution. Hence, there is not any memory effect at the level of single
realizations. Alternatively, a non-Markovian wave vector evolution that in
average recovers the system evolution may also be proposed \cite{piiloSWF}.
These examples confirm that a procedure capable to determine when memory
effects rely or not on physically mediated BIFs is in general highly
demanded.

The aim of this work is introduce an operational technique that accurately
detects the presence of physically mediated system-environment BIFs.
Consistently with the operational character, instead of a definition in the
system Hilbert space \cite{BreuerReview,plenioReview}, the approach relies
on a probabilistic condition that indicates when an environment is
unaffected by its coupling with the system. Correspondingly, memory effects
emerge from a statistical average of a Markovian system dynamics that
parametrically depends\ on the (unaffected) bath degrees of freedom. It is
shown that these conditions can be checked by performing a minimal number of
three system measurement processes, added to an intermediate (random) update
of the system state that may depends on previous outcomes. Similarly to
operational memory approaches based on causal breaks \cite%
{modi,budiniCPF,pollock,pollockInfluence,bonifacio,budiniChina,budiniBrasil,han,goan}%
, here a generalized conditional past-future (CPF) correlation \cite%
{budiniCPF,budiniChina,budiniBrasil,bonifacio,han} defined between the first
and last (past-future) measurement outcomes, conditioned to the intermediate
updated system-state, becomes an indicator of BIFs.

The three-joint outcome probabilities and its associated generalized CPF
correlation are calculated for both quantum and classical environmental
fluctuations. Consistently, for classical noise fluctuations, or in general,
when memory effects can be associated to environments with an invariant
dynamics, the generalized CPF correlation vanishes. This property furnishes
a novel and explicit experimental test for detecting BIFs. Its feasibility
is explicitly demonstrated through its characterization in ubiquitous
dissipative and dephasing non-Markovian dynamics that admit an exact
treatment.

\section{Probabilistic approach}

Our aim is to distinguish between memory effects that occur with and without
BIFs. These opposite cases are related to the dependence or independence of
the reservoir dynamics on system degrees of freedom. This property can be
explicitly defined by means of the following scheme, which is valid in both
classical and quantum realms.

We assume that both the system and the environment are subjected to a set of
(bipartite separable) measurements at successive times $t_{1}\!<\!t_{2}%
\cdots \!<\!t_{n}.$ The set of strings $\mathbf{s}\equiv (s_{1},s_{2}\cdots
s_{n})$ and $\mathbf{e}\equiv (e_{1},e_{2},\cdots e_{n})$ denote the
respective outcomes, which in turn label the corresponding system and
environment post-measurement states. The outcome statistics is set by a
joint probability $P(\mathbf{s},\mathbf{e}).$ This object in general depends
on which measurement processes are performed.

In agreement with our definition, in \textit{absence} of BIFs the
environment probability $P\mathbf{(e)}=\sum_{\mathbf{s}}P(\mathbf{s},\mathbf{%
e})$ must be an invariant object that is independent of the system
initialization and dynamics. Bayes rule allows to write $P\mathbf{(e)}=\sum_{%
\mathbf{s}}P(\mathbf{e|s})P(\mathbf{s}),$ where $P(\mathbf{e|s})$ is the
conditional probability of $\mathbf{e}$ given $\mathbf{s},$ while $P(\mathbf{%
s})$ gives the probability of $\mathbf{s.}$ Hence, the absence of BIFs can
be expressed by the condition%
\begin{equation}
P(\mathbf{e|s})=P(\mathbf{e}),  \label{Zero}
\end{equation}%
which guarantees that the environment statistics is independent of the
system state and dynamics.

The marginal probability for the system outcomes can always be written as $P%
\mathbf{(s)}=\sum_{\mathbf{e}}P(\mathbf{s},\mathbf{e})=\sum_{\mathbf{e}}P(%
\mathbf{s|e})P(\mathbf{e}),$ where $P(\mathbf{s|e})$ is the conditional
probability of $\mathbf{s}$ given $\mathbf{e}.$ When condition (\ref{Zero})
is fulfilled, we can affirm that any possible memory effect in the system
measurements follows from an (invariant) environmental average $[\langle
\cdots \rangle _{\mathbf{e}}\mathbf{\equiv }\sum\nolimits_{\mathbf{e}}\cdots
P\mathbf{(e)}]$ of a (system) joint probability $P^{(\mathbf{e})}(\mathbf{s}%
)\leftrightarrow P(\mathbf{s|e})$ that parametrically depends on the bath
states,%
\begin{equation}
P(\mathbf{s})=\langle P^{(\mathbf{e})}(\mathbf{s})\rangle _{\mathbf{e}}.
\label{definittionBIF}
\end{equation}%
Notice that $P^{(\mathbf{e})}(\mathbf{s})$ denotes the conditional
probability $P(\mathbf{s|e})$ given that condition (\ref{Zero}) is fulfilled.

In the present approach Eqs. (\ref{Zero}) and (\ref{definittionBIF}) \textit{%
define} the absence of any physical system-environment BIF. System memory
effects emerge due to the conditional action of the bath. Our problem now is
to detect these probability structures by taking into account only the
system outcome statistics. Before this step, we introduce one extra
assumption.

As usual in open quantum systems, we assume that the system-bath bipartite
dynamics (without interventions) admits an underlying semigroup (memoryless)
description. Hence, $P^{(\mathbf{e})}(\mathbf{s})$ fulfills a Markovian
property with respect to system outcomes,%
\begin{equation}
P^{(\mathbf{e})}(\mathbf{s})=P^{(\mathbf{e})}(s_{n}|s_{n-1})\cdots P^{(%
\mathbf{e)}}(s_{2}|s_{1})P^{(\mathbf{e})}(s_{1}).  \label{MarkovHypothesis}
\end{equation}%
For notational convenience, the parametric dependence of the conditional
probabilities $P^{(\mathbf{e)}}(s|s^{\prime })$\ on the bath states is
written through the supra index $(\mathbf{e)}.$ This dependence must be
consistent with causality, meaning that $P^{(\mathbf{e)}}(s|s^{\prime })$
cannot depend on (non-selected) future bath outcomes.

\subsection{Detection scheme}

The developing of BIFs, that is, departures with respect to the structure
defined by Eqs.~(\ref{definittionBIF})-(\ref{MarkovHypothesis}), can be
detected with the following minimal scheme. Three measurements processes
performed at times $0\rightarrow t\rightarrow t+$ $\tau ,$ deliver the
successive system outcomes $x~\rightarrow ~(y\rightarrow ~\breve{y}%
)\rightarrow z.$ After the intermediate measurement, the system
state---labelled by $y$---is \textit{externally} (and instantaneously)
updated to a renewed state---labelled by $\breve{y}$---, while the bath
state is unaffected. Each $\breve{y}$-state is chosen with an arbitrary
conditional probability $\wp (\breve{y}|y,x).$ The scheme is closed after
specifying $\wp (\breve{y}|y,x)$ and calculating the marginal probability $%
P(z,\breve{y},x)=\sum_{y}P(z,\breve{y},y,x).$ In addition, it is assumed
that system and environment are uncorrelated before the first measurement. A
\textquotedblleft \textit{deterministic scheme}\textquotedblright\ (d)
corresponds to\ $\wp (\breve{y}|y,x)=\delta _{\breve{y},y}.$ Hence, not any
change is introduced after the intermediate measurement. A \textquotedblleft 
\textit{random scheme}\textquotedblright\ (r) is defined by $\wp (\breve{y}%
|y,x)=\wp (\breve{y}|x).$ These two cases are motivated by the following
features.

In \textit{absence} of BIFs, the joint probability for the four events, from
Eqs.~(\ref{definittionBIF}) and (\ref{MarkovHypothesis}), reads%
\begin{equation}
P(z,\breve{y},y,x)=\langle P^{(\mathbf{e})}(z|\breve{y})\wp (\breve{y}%
|y,x)P^{(\mathbf{e})}(y|x)P(x)\rangle _{\mathbf{e}}.  \label{prob-four}
\end{equation}%
Notice that this result also relies on Eq. (\ref{Zero}), which guarantees
that $\langle \cdots \rangle _{\mathbf{e}}$ remains invariant even when
changing the system state at a given time, $(y\rightarrow ~\breve{y}).$ On
the other hand, by assumption $\wp (\breve{y}|y,x)$ and $P(x)$ do not depend
on the environmental degrees of freedom. In the deterministic scheme, Eq.~(%
\ref{prob-four}) leads to%
\begin{equation}
P(z,\breve{y},x)\overset{d}{=}\langle P^{(\mathbf{e})}(z|\breve{y})P^{(%
\mathbf{e})}(\breve{y}|x)\rangle _{\mathbf{e}}\,P(x),  \label{MemorySinBIF}
\end{equation}%
while in the random case, using $\sum_{y}P^{(\mathbf{e})}(y|x)=1,$%
\begin{equation}
P(z,\breve{y},x)\overset{r}{=}\langle P^{(\mathbf{e})}(z|\breve{y})\rangle _{%
\mathbf{e}}\,\wp (\breve{y}|x)P(x).  \label{MarkovInducido}
\end{equation}

The deterministic scheme [Eq.~(\ref{MemorySinBIF})], given that $P(z,\breve{y%
},x)$ does not fulfill a Markov property, shows that memory effects may in
fact develop even in absence of BIFs. Nevertheless, due to the structure
defined by Eqs.~(\ref{definittionBIF}) and (\ref{MarkovHypothesis}), they
are completely \textquotedblleft washed out\textquotedblright\ in the random
scheme, which delivers a Markovian joint probability [Eq.~(\ref%
{MarkovInducido})]. Taking into account the derivation of Eq.~(\ref%
{prob-four}), this last property break down when Eq.~(\ref{Zero}) is not
fulfilled. Thus, \textit{in the random scheme departure of }$P(z,\breve{y}%
,x) $\textit{\ from Markovianity witnesses BIFs, }which solves our problem.

\subsection{System and environment observables}

In contrast to classical systems, in a quantum regime the previous results
have an intrinsic dependence of which system and environment observables are
considered.

For quantum systems, the \textit{absence of BIFs} is defined by the validity
of the probability structures Eqs.~(\ref{MemorySinBIF}) and (\ref%
{MarkovInducido}) for any kind of system measurement processes. Thus, 
\textit{arbitrary system observables} are considered.

On the other hand, we \textit{only} consider environment observables that
allow to read $\langle \cdots \rangle _{\mathbf{e}}$ as an \textit{%
unconditional }average over the bath degrees of freedom. This extra
assumption is completely consistent with the developed approach.
Furthermore, this election (due to the unconditional character)\ implies
that $P(z,\breve{y},x)$ can be measured without involving any \textit{%
explicit} environment measurement process. This important feature is valid
for both classical and quantum environmental fluctuations.

When the environment is defined by classical stochastic degrees of freedom
with a fixed statistics [Sec.~(\ref{clasicor})], given that classical
systems are not affected by a measurement process, the previous assumption
applies straightforwardly. When the reservoir must be described in a quantum
regime, the previous constraint implies observables whose \textit{%
non-selective}~\cite{breuerbook} measurement transformations do not affect
the environment state at each stage [Sec.~(\ref{quantor})]. Thus,
independently of the environment nature, the detection of BIFs can always be
performed without measuring explicitly the environment.

\subsection{BIF witness}

Independently of the nature (incoherent or quantum) of both the system and
the environment, as an explicit witness of BIF we consider a generalized CPF
correlation that takes into account the intermediate system state update
operation (deterministic $\leftrightarrow d$ or random $\leftrightarrow r$).
It measures the correlation between the initial and final (past-future)
outcomes conditioned to the intermediate system state ($\breve{y}$)%
\begin{equation}
C_{pf}^{(d/r)}|_{\breve{y}}\equiv \sum_{z,x}O_{z}O_{x}[P(z,x|\breve{y})-P(z|%
\breve{y})P(x|\breve{y})].  \label{GCPF}
\end{equation}%
Here, all conditional probabilities follow from $P(z,\breve{y},x)$ \cite%
{Conditionals}, while the sum indexes run over all possible outcomes at each
stage. The scalar quantities $\{O_{z}\}$ and $\{O_{x}\}$ define the system
observables for each outcome.

In the deterministic scheme, similarly to Ref.$~$\cite{budiniCPF}, $%
C_{pf}^{(d)}|_{\breve{y}}$ detects memory effects independently of its
underlying origin. In the random scheme, the condition $C_{pf}^{(r)}|_{%
\breve{y}}~\neq ~0$ provides the desired witness of BIFs. This result
follows directly from the Markovian property Eq.~(\ref{MarkovInducido}),
which leads to $P(z,x|\breve{y})=P(z|\breve{y})P(x|\breve{y})\rightarrow
C_{pf}^{(r)}|_{\breve{y}}=0.$

For quantum systems, the three system measurement processes are defined by a
set of operators $\{\Omega _{x}\},$ $\{\Omega _{y}\},$ and $\{\Omega _{z}\},$
with normalization $\sum\nolimits_{x}\Omega _{x}^{\dag }\Omega
_{x}=\sum\nolimits_{y}\Omega _{y}^{\dag }\Omega _{y}=\sum\nolimits_{z}\Omega
_{z}^{\dag }\Omega _{z}=\mathrm{I,}$ where $\mathrm{I}$ is the system
identity operator. The intermediate $y$-measurement in taken as a projective
one, $\Omega _{y}=|y\rangle \langle y|.$ Thus, in the random scheme the
system state transformation reads $\rho _{y}\equiv |y\rangle \langle
y|\rightarrow \rho _{\breve{y}},$ where the states $\{\rho _{\breve{y}}\}$
(independently of outcome $y)$ are randomly chosen with probability $\wp (%
\breve{y}|x).$ This operation can be implemented, for example, as $\rho _{%
\breve{y}}=U(\breve{y}|y)[\rho _{y}],$ where the (conditional) unitary
operator $U(\breve{y}|y)$ leads to the state $\rho _{\breve{y}}$
independently of the obtained $y$-outcome \cite{repraration}.

\section{Application to different system-environment models}

The consistence of the developed approach is supported by studying
fundamental system-reservoir models that leads to memory effects.

\subsection{Classical noise environmental fluctuations}

\label{clasicor}

Here the open system is coupled to classical stochastic degrees of freedom.
Its density matrix is written as $\rho _{t}=\overline{\mathcal{E}_{t,0}^{st}}%
[\rho _{0}],$ where the overbar symbol denotes an average over the
environmental realizations. For each noise realization the stochastic
propagator fulfills $\mathcal{E}_{t+\tau ,0}^{st}=\mathcal{E}_{t+\tau
,t}^{st}\mathcal{E}_{t,0}^{st},$ property consistent with the assumption (%
\ref{MarkovHypothesis}). Stochastic Hamiltonians \cite%
{cialdi,GaussianNoise,morgado,bordone} as well as random unitary evolutions 
\cite{wudarski} fall in this category. As usual in these models, the
statistics of the noise realizations is independent of the system dynamics.
Hence, not any BIF should be detected in this case.

Given that each noise realization labels the environment state, we can take
the equivalence $\langle \cdots \rangle _{\mathbf{e}}\leftrightarrow 
\overline{(\cdots )}.$ By using the standard formulation of quantum
measurement theory, the joint probability associated to the measurement
scheme can be written as (see Appendix~A)%
\begin{equation}
\frac{P(z,\breve{y},y,x)}{\wp (\breve{y}|y,x)}=\overline{\mathrm{Tr}%
_{s}(E_{z}\mathcal{E}_{t+\tau ,t}^{st}[\rho _{\breve{y}}])\mathrm{Tr}%
_{s}(E_{y}\mathcal{E}_{t,0}^{st}[\tilde{\rho}_{x}])},
\label{PJoint4StochHamiltonian}
\end{equation}%
where $E_{i}\equiv \Omega _{i}^{\dagger }\Omega _{i}$ $(i=x,y,z)$ and $%
\tilde{\rho}_{x}\equiv \Omega _{x}\rho _{0}\Omega _{x}^{\dagger }$ is the
(unnormalized) system state after the first $x$-measurement. $\mathrm{Tr}%
_{s}(\cdots )$ denotes a trace operation in the system Hilbert space. $\rho
_{\breve{y}}$ is the (updated) system state after the second $y$%
-measurement, while $t$ and $\tau $ are the elapsed times between
consecutive measurements.

In the deterministic scheme $[\wp (\breve{y}|y,x)=\delta _{\breve{y},y}],$
using that $P(z,\breve{y},x)=\sum_{y}P(z,\breve{y},y,x),$ Eq. (\ref%
{PJoint4StochHamiltonian}) leads to%
\begin{equation}
P(z,\breve{y},x)\overset{d}{=}\overline{\mathrm{Tr}_{s}(E_{z}\mathcal{E}%
_{t+\tau ,t}^{st}[\rho _{\breve{y}}])\mathrm{Tr}_{s}(E_{\breve{y}}\mathcal{E}%
_{t,0}^{st}[\tilde{\rho}_{x}])}.  \label{PZXStochastic}
\end{equation}%
In general, this joint probability does not fulfill a Markov condition.
Thus, $C_{pf}^{(d)}|_{\breve{y}}\neq 0$ [Eq.~(\ref{GCPF})] detects memory
effects. On the other hand, in the random scheme $[\wp (\breve{y}|y,x)=\wp (%
\breve{y}|x)]$ from Eq.~(\ref{PJoint4StochHamiltonian}) it follows%
\begin{equation}
P(z,\breve{y},x)\overset{r}{=}\overline{\mathrm{Tr}_{s}(E_{z}\mathcal{E}%
_{t+\tau ,t}^{st}[\rho _{\breve{y}}])}\wp (\breve{y}|x)\mathrm{Tr}_{s}(%
\tilde{\rho}_{x}),  \label{ProbRandomNoise}
\end{equation}%
which recovers the Markovian result Eq.~(\ref{MarkovInducido}) with $\langle
P^{(\mathbf{e})}(z|\breve{y})\rangle _{\mathbf{e}}\leftrightarrow \overline{%
\mathrm{Tr}_{s}(E_{z}\mathcal{E}_{t+\tau ,t}^{st}[\rho _{\breve{y}}])}=P(z|%
\breve{y})$ and $P(x)=\mathrm{Tr}_{s}(\tilde{\rho}_{x})=\mathrm{Tr}%
_{s}(E_{x}\rho _{0}).$ Thus, \textit{independently of the chosen system
measurement observables it follows }$C_{pf}^{(r)}|_{\breve{y}}~=~0$ [Eq.~(%
\ref{GCPF})], indicating, as expected, the absence of any BIF.

\subsection{Completely positive system-environment dynamics}

\label{quantor}

Alternatively, system-environment ($s$-$e$) dynamics can be described in a
bipartite Hilbert space. Their density matrix $\rho _{t}^{se}=\mathcal{E}%
_{t,0}[\rho _{0}^{se}]$ is set by a bipartite propagator that satisfies $%
\mathcal{E}_{t+\tau ,0}=\mathcal{E}_{t+\tau ,t}\mathcal{E}_{t,0}.$ This
property also supports assumption (\ref{MarkovHypothesis}). We consider
separable initial conditions $\rho _{0}^{se}=\rho _{0}\otimes \sigma _{0}.$
Hence, $\mathcal{E}_{t,0}$ leads to a completely positive system dynamics $%
\rho _{t}=\mathrm{Tr}_{e}(\mathcal{E}_{t,0}[\rho _{0}^{se}]).$ Unitary
system-environment models \cite{breuerbook} as well as bipartite
(time-irreversible) Lindblad dynamics\ fall in this category. As system and
environment are intrinsically coupled, the developing of BIFs is expected in
general.

Here, we take the equivalence $\langle \cdots \rangle _{\mathbf{e}%
}\leftrightarrow \mathrm{Tr}_{e}(\cdots ).$ This \textit{unconditional
environment average} applies when the successive (non-selective \cite%
{breuerbook}) measurements of the environment do not modify its state at
each stage (the bath state remains the same after each non-selective
measurement). Due to the dynamics induced by $\mathcal{E}_{t,0},$ in general
it is not possible to know explicitly\ which physical reservoir observables
fulfill this condition. Nevertheless,\ the demanded invariance
straightforwardly allows to read and to obtain $\langle \cdots \rangle _{%
\mathbf{e}}$ from the bath trace operation $\mathrm{Tr}_{e}(\cdots )$ \cite%
{breuerbook}\ (see also Appendix~A). Hence, similarly to the previous
environment model the validity (or not) of Eqs. (\ref{MemorySinBIF}) and (%
\ref{MarkovInducido}) can be checked without performing any \textit{explicit}
reservoir measurement process. From standard quantum measurement theory, the
joint probability of system outcomes here reads (Appendix~A)%
\begin{equation}
\frac{P(z,\breve{y},y,x)}{\wp (\breve{y}|y,x)}=\mathrm{Tr}_{se}(E_{z}%
\mathcal{E}_{t+\tau ,t}[\rho _{\breve{y}}\otimes \mathrm{Tr}_{s}(E_{y}%
\mathcal{E}_{t,0}[\tilde{\rho}_{x}^{se}])]),  \label{Prob4CP}
\end{equation}%
where $\tilde{\rho}_{x}^{se}\equiv \Omega _{x}\rho _{0}\Omega _{x}^{\dagger
}\otimes \sigma _{0}=\tilde{\rho}_{x}\otimes \sigma _{0}$ is the bipartite
state after the\ first $x$-measurement and, as before, $\rho _{\breve{y}}$
is the updated system state.

In the deterministic scheme $[\wp (\breve{y}|y,x)=\delta _{\breve{y},y}],$
the previous expression $[P(z,\breve{y},x)=\sum_{y}P(z,\breve{y},y,x)]$
leads to%
\begin{equation}
P(z,\breve{y},x)\overset{d}{=}\mathrm{Tr}_{se}(E_{z}\mathcal{E}_{t+\tau
,t}[\rho _{\breve{y}}\otimes \mathrm{Tr}_{s}(E_{\breve{y}}\mathcal{E}_{t,0}[%
\tilde{\rho}_{x}^{se}])]).  \label{DeterministicBipartito}
\end{equation}%
As expected, a Markovian property is not fulfilled in general implying the
presence of memory effects, $C_{pf}^{(d)}|_{\breve{y}}\neq 0.$ In the random
scheme $[\wp (\breve{y}|y,x)=\wp (\breve{y}|x)]$ it follows%
\begin{equation}
P(z,\breve{y},x)\overset{r}{=}\mathrm{Tr}_{se}(E_{z}\mathcal{E}_{t+\tau
,t}[\rho _{\breve{y}}\otimes \mathrm{Tr}_{s}(\mathcal{E}_{t,0}[\tilde{\rho}%
_{x}^{se}])])\wp (\breve{y}|x).  \label{RandomBipartito}
\end{equation}%
In contrast to Eq.~(\ref{ProbRandomNoise}), here in general a Markov
property is not fulfilled. Thus, $C_{pf}^{(r)}|_{\breve{y}}\neq 0.$
Nevertheless, there are bipartite dynamics than in fact occur without a BIF.
Below, we found the conditions that guarantee $C_{pf}^{(r)}|_{\breve{y}}=0$
for arbitrary system measurement processes.

\subsubsection{Invariant environment dynamics}

The environment state follows by tracing out the system degrees of freedom, $%
\sigma _{t}\equiv \mathrm{Tr}_{s}(\mathcal{E}_{t,0}[\rho _{0}^{se}]),$ where 
$\rho _{0}^{se}=\rho _{0}\otimes \sigma _{0}.$ When this state is
independent of the system initialization%
\begin{equation}
\sigma _{t}=\mathrm{Tr}_{s}(\mathcal{E}_{t,0}[\rho _{0}^{se}])=\mathrm{Tr}%
_{s}(\mathcal{E}_{t,0}[\mathcal{M}_{s}[\rho _{0}^{se}]]),
\label{Nonsignalin}
\end{equation}%
where $\mathcal{M}_{s}$ represents an arbitrary (trace-preserving) system
transformation, a Markovian property is immediately recovered in the random
scheme. In fact, introducing $\mathrm{Tr}_{s}(\mathcal{E}_{t,0}[\tilde{\rho}%
_{x}^{se}])])=P(x)\sigma _{t},$ Eq.~(\ref{RandomBipartito}) becomes $P(z,%
\breve{y},x)\overset{r}{=}\mathrm{Tr}_{se}(E_{z}\mathcal{E}_{t+\tau ,t}[\rho
_{\breve{y}}\otimes \sigma _{t}])\wp (\breve{y}|x)P(x),$ which recovers the
structure (\ref{MarkovInducido}). Thus,\textit{\ environments with an
invariant dynamics do not induce any BIF} $[C_{pf}^{(r)}|_{\breve{y}}=0].$
Notice that this property supports the complete consistence of the proposed
approach.

A relevant situation where Eq.~(\ref{Nonsignalin}) applies is the case of
systems coupled to incoherent degrees of freedom governed by a (invariant)
classical master equation \cite{lindbladrate}. While these dynamics lead to
memory effects \cite{PostMarkovian,shabani,boltzman}, our approach correctly
identify the absence of any BIF. Random unitary evolutions \cite{wudarski},\
as well as quantum Markov chains \cite{megier,maximal} fall in this case.

It is important to remark that environments developing quantum features
(coherences) may also fulfill condition (\ref{Nonsignalin}). This is the
case, for example, of some collisional models \cite{colisionVacchini} whose
underlying description can be formulated with bipartite Lindblad equations 
\cite{embedding}.

\subsubsection{Unitary system-environment models}

When modeling open quantum dynamics from an underlying bipartite Hamiltonian
dynamics, the unitary propagator reads $\mathcal{E}_{t,0}[\cdot ]=\exp
(-itH_{T})\cdot \exp (+itH_{T}),$ where $H_{T}$ is%
\begin{equation}
H_{T}=H_{s}+H_{e}+H_{I}.  \label{Htotal}
\end{equation}%
The first two terms define respectively the system and bath Hamiltonians,
while the last one introduces their interaction. Given the
system-environment mutual interaction, for nearly all Hamiltonians $H_{T}$
it is expected that the developing of memory effects [Eq.~(\ref%
{DeterministicBipartito})] rely on BIFs [Eq.~(\ref{RandomBipartito})].

One exception to the previous rule arises when the bath and interaction
Hamiltonians commute,%
\begin{equation}
\lbrack H_{e},H_{I}]=0.  \label{Conmutador}
\end{equation}%
Under this condition, denoting the bath eigenvectors as $H_{e}|e\rangle
=e|e\rangle ,$ the system density matrix reads $\rho _{t}=\mathrm{Tr}%
_{e}(\rho _{t}^{se})=\sum\nolimits_{e}w_{e}\exp (-itH_{s}^{(e)})\rho
_{0}\exp (+itH_{s}^{(e)}),$ where the weights are $w_{e}\equiv \langle
e|\sigma _{0}|e\rangle $ and $H_{s}^{(e)}\equiv H_{s}+\langle
e|H_{I}|e\rangle .$ Thus, the system dynamics can be represented by a random
unitary map \cite{nori}. For arbitrary dynamics, this property does not
guaranty the absence of BIFs. In fact, here the environment invariance
property~(\ref{Nonsignalin}) is not fulfilled in general \cite{invariance}.
Nevertheless, after a straightforward calculation, the probabilities of the
deterministic and random schemes, Eqs.~(\ref{DeterministicBipartito}) and (%
\ref{RandomBipartito}), can be written as in Eqs.~(\ref{MemorySinBIF}) and (%
\ref{MarkovInducido}) (valid in absence of BIFs) respectively. In fact,
under the replacement $\langle \cdots \rangle _{\mathbf{e}}\rightarrow
\sum\nolimits_{e}w_{e}(\cdots ),$ the conditional probabilities are $P^{(%
\mathbf{e})}(z|\breve{y})\rightarrow \mathrm{Tr}_{s}(E_{z}\mathbb{G}_{\tau
}^{(e)}[\rho _{\breve{y}}])$ and $P^{(\mathbf{e})}(\breve{y}|x)\rightarrow 
\mathrm{Tr}_{s}(E_{\breve{y}}\mathbb{G}_{t}^{(e)}[\rho _{x}]),$ where $%
\mathbb{G}_{t}^{(e)}[\cdot ]\equiv \exp (-itH_{s}^{(e)})\cdot \exp
(+itH_{s}^{(e)})$ and $\rho _{x}\equiv \tilde{\rho}_{x}/\mathrm{Tr}_{s}(%
\tilde{\rho}_{x}).$ Thus, from these expressions we conclude that the
condition (\ref{Conmutador}) guaranties that the joint probabilities, for
arbitrary system measurement processes, can also be obtained from a
statistical mixture (with invariant weights $\{w_{e}\}$) of unitary system
evolutions (with propagators $\{\mathbb{G}_{t}^{(e)}\}$), which consistently
implies $C_{pf}^{(r)}|_{\breve{y}}=0.$

\section{Examples}

Here, different explicit examples that admit an exact treatment are studied.

\subsection{Eternal non-Markovianity}

As a first explicit example we consider the non-Markovian system evolution%
\begin{equation}
\frac{d\rho _{t}}{dt}=\frac{1}{2}\sum_{\alpha =\hat{x},\hat{y},\hat{z}%
}\gamma _{\alpha }(t)(\sigma _{\alpha }\rho _{t}\sigma _{\alpha }-\rho _{t}),
\label{MasterEternal}
\end{equation}%
where $\{\sigma _{\alpha }\}$ are the $\alpha $-Pauli matrixes (directions
in Bloch sphere are denoted with a hat symbol). The time-dependent rates\
are $\gamma _{\hat{x}}(t)=\gamma _{\hat{y}}(t)=\gamma ,$ and $\gamma _{\hat{z%
}}(t)=-\gamma \tanh [\gamma t].$ As demonstrated in Ref.~\cite{megier} this
kind of \textit{eternal} non-Markovian evolution $[\gamma _{\hat{z}}(t)<0$ $%
\forall t]$ is induced by the coupling of the system with a statistical
mixture of classical random fields. In fact, the system state can be written
as $\rho _{t}=\sum_{\alpha =\hat{x},\hat{y},\hat{z}}q_{\alpha }\exp [\gamma t%
\mathbb{L}_{\alpha }][\rho _{0}],$ where $\mathbb{L}_{\alpha }[\cdot ]\equiv
(\sigma _{\alpha }\cdot \sigma _{\alpha }-\cdot )$ is induced by each random
field, whose (mixture) weights are $q_{\hat{x}}=q_{\hat{y}}=1/2,$ and $q_{%
\hat{z}}=0.$ This underlying \textquotedblleft
microscopic\textquotedblright\ description allows to calculating multi-time
statistics in an exact way. In particular, the CPF correlations follow
straightforwardly from Eqs.~(\ref{PZXStochastic}) and (\ref{ProbRandomNoise}%
),\ $\overline{(\cdots )}\rightarrow \sum_{\alpha =\hat{x},\hat{y},\hat{z}%
}q_{\alpha }(\cdots ),$ where the (time-independent) \textquotedblleft noise
environmental realizations\textquotedblright\ only assumes the values $%
\alpha =\hat{x},\hat{y},\hat{z},$ each with probability $q_{\alpha }.$

Assuming that the three measurements processes are performed in the Bloch
directions $\hat{x}$-$\hat{n}$-$\hat{x},$ where $\hat{n}$ is an arbitrary
direction in the $\hat{z}$-$\hat{x}$\ plane (with azimuthal angle $\theta $%
), for the deterministic scheme it follows (see Appendix~B)%
\begin{equation}
C_{pf}^{(d)}|_{\breve{y}=\pm 1}\underset{\hat{x}\hat{n}\hat{x}}{=}\sin
^{2}(\theta )[c(t+\tau )-c(t)c(\tau )],  \label{CPFDetEternal}
\end{equation}%
where $c(t)\equiv q_{\hat{x}}+(q_{\hat{y}}+q_{\hat{z}})\exp [-2\gamma t].$
The initial system state was taken as $\rho _{0}=|\pm \rangle \langle \pm |,$
where $|\pm \rangle $ denotes the eigenvectors of $\sigma _{\hat{z}}.$ In
Fig. 1(a) we plot $C_{pf}^{(d)}|_{\breve{y}}$ [Eq.~(\ref{CPFDetEternal})]
and $C_{pf}^{(r)}|_{\breve{y}}$ for equal measurement time intervals, $%
t=\tau .$ The property $\lim_{t\rightarrow \infty }C_{pf}^{(d)}|_{\breve{y}%
}\neq 0$ indicates that the environment correlation do not decay in time 
\cite{budiniCPF}. On the other hand, independently of the election of the
renewed (pure) states $\rho _{\breve{y}=\pm 1}$ and $\wp (\breve{y}|x),$ we
get $C_{pf}^{(r)}|_{\breve{y}}=0$ (see Appendix B). As expected from Eq.~(%
\ref{ProbRandomNoise}), this result indicates the absence of any BIF.

\subsection{Interaction with a bosonic bath}

As a second example, we consider a two-level system coupled to a bosonic
bath,%
\begin{equation}
H_{T}=\frac{\omega _{0}}{2}\sigma _{\hat{z}}+\sum_{k}\omega _{k}b_{k}^{\dag
}b_{k}+\sum_{k}g_{k}Sb_{k}^{\dag }+g_{k}^{\ast }S^{\dagger }b_{k}.
\label{Bosonico}
\end{equation}%
Each contribution defines the system, bath, and interaction Hamiltonians
respectively [Eq. (\ref{Htotal})]. The bosonic operators satisfy $%
[b_{k},b_{k^{\prime }}^{\dag }]=\delta _{k,k^{\prime }}.$ Taking the system
operators $S^{\dagger }=\left\vert {+}\right\rangle \left\langle {-}%
\right\vert $ and $S=\left\vert {-}\right\rangle \left\langle {+}\right\vert 
$ as the raising and lowering operators in the natural basis $\left\vert {%
\pm }\right\rangle ,$ the system dynamics is dissipative \cite{breuerbook},
while in the case $S=S^{\dagger }=\sigma _{\hat{z}}$ a dephasing dynamics is
recovered. We assume the bipartite initial state $|\Psi _{0}^{se}\rangle
=|\psi _{0}\rangle \otimes \prod_{k}|0\rangle _{k},$ where $\{|0\rangle
_{k}\}$ are the ground states of each bosonic mode. In this case, by working
the observables in an interaction representation, similarly to Refs. \cite%
{budiniChina,budiniBrasil}, the joint probabilities (\ref%
{DeterministicBipartito}) and (\ref{RandomBipartito}) can be calculated in
an exact way \cite{unpublished}. 
\begin{figure}[tbp]
\includegraphics[bb=25 220 515 795,angle=0,width=8.5cm]{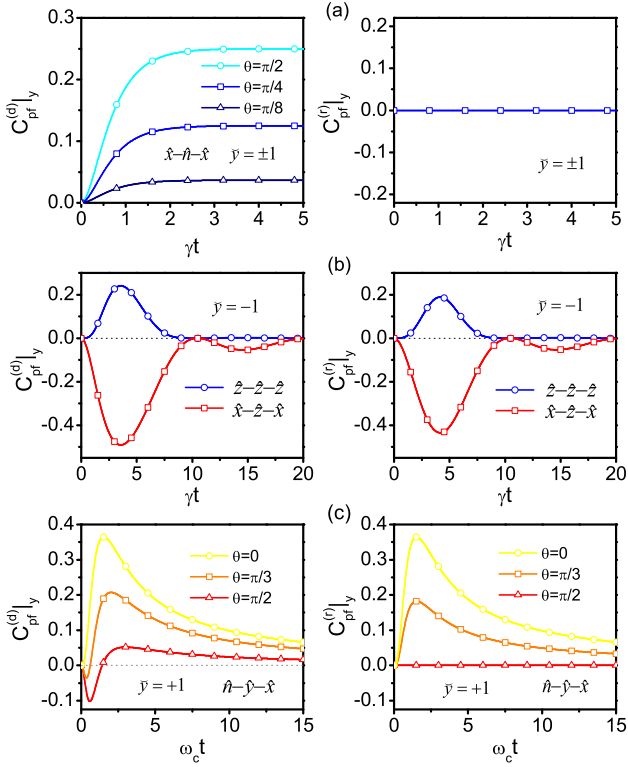}
\caption{CPF correlation [Eq.~(\protect\ref{GCPF})] for the deterministic
and random schemes, left and right columns respectively, for equal
measurement time intervals $t=\protect\tau .$ (a) Eternal non-Markovianity,
measurements $\hat{x}$-$\hat{n}$-$\hat{x}.$ (b) Decay in a bosonic bath,
measurements $\hat{z}$-$\hat{z}$-$\hat{z}$ and $\hat{x}$-$\hat{z}$-$\hat{x}.$
(c) Dephasing in a bosonic bath, measurements $\hat{n}$-$\hat{y}$-$\hat{x}.$
In all cases, the $\hat{n}-$direction is defined by the angle $\protect%
\theta .$ The renewed states $\protect\rho _{\breve{y}=\pm 1}$ are described
in the main text.}
\end{figure}

For the \textit{dissipative }dynamics [$S=\left\vert {-}\right\rangle
\left\langle {+}\right\vert $ in Eq.~(\ref{Bosonico})] the CPF correlation
in the random scheme reads \cite{unpublished}%
\begin{equation}
C_{pf}^{(r)}|_{\breve{y}=-1}\underset{\hat{z}\hat{z}\hat{z}}{=}|G(t,\tau
)|^{2},\ \ \ \ \ C_{pf}^{(r)}|_{\breve{y}=-1}\underset{\hat{x}\hat{z}\hat{x}}%
{=}-\mathrm{Re}[G(t,\tau )].
\end{equation}%
Here, we consider two different measurement possibilities, $\hat{z}$-$\hat{z}
$-$\hat{z}$ and $\hat{x}$-$\hat{z}$-$\hat{x}$ directions, both with
conditional $\breve{y}=-1.$ The renewed states are $\rho _{\breve{y}=\pm
}=|\pm \rangle \langle \pm |,$ and we take $\wp (\breve{y}|x)=1/2.$ The
initial system state $|\psi _{0}\rangle $ is chosen such that $P(x)=1/2.$
Under this condition, for both measurement directions, in the deterministic
scheme we get $C_{pf}^{(d)}|_{\breve{y}%
=-1}=[1-|G(t)|^{2}/2]^{-2}C_{pf}^{(r)}|_{\breve{y}=-1}.$ In these
expressions, $G(t,\tau )\equiv \int_{0}^{t}dt^{\prime }\int_{0}^{\tau }d\tau
^{\prime }f(\tau ^{\prime }+t^{\prime })G(t-t^{\prime })G(\tau -\tau
^{\prime }),$ where $G(t)$ is defined by the evolution$\
(d/dt)G(t)=-\int_{0}^{t}f(t-t^{\prime })G(t^{\prime })dt^{\prime },$ $%
G(0)=1. $ The memory kernel is the bath correlation $f(t)\equiv
\sum_{k}|g_{k}|^{2}\exp [+i(\omega _{0}-\omega _{k})t].$

In Fig. 1(b), for a Lorentzian spectral density \cite{budiniBrasil}, $%
f(t)=(\gamma /2\tau _{c})\exp (-|t|/\tau _{c}),$ with $\gamma \tau _{c}=5,$
we plot the CPF correlations. In contrast to the previous case, here for
both the deterministic and random schemes, the CPF correlations do not
vanish. Thus, memory effects rely on BIFs, which are present independently
of the bath correlation time $\tau _{c}.$

In the \textit{dephasing} case [$S=\sigma _{\hat{z}}$ in Eq.~(\ref{Bosonico}%
)], the CPF correlation in the random scheme is \cite{unpublished}%
\begin{equation}
C_{pf}^{(r)}|_{\breve{y}}\underset{\hat{n}\hat{y}\hat{x}}{=}\breve{y}\cos
(\theta )\exp (-\gamma _{\tau })\sin (\Phi _{t,\tau }).
\end{equation}%
Here, we consider the successive measurements in Bloch directions $\hat{n}$-$%
\hat{y}$-$\hat{x}.$ Furthermore, we take $\wp (\breve{y}|x)=1/2,$ and pure
states $\rho _{\breve{y}}$ corresponding to the eigenvectors of $\sigma _{%
\hat{y}}.$ The initial condition $|\psi _{0}\rangle $ is such that
independently of $\hat{n},$ $P(x)=1/2.$ Under this condition the CPF
correlation of the deterministic scheme can be written as $C_{pf}^{(d)}|_{%
\breve{y}}\underset{\hat{n}\hat{y}\hat{x}}{=}\sin (\theta )\exp [-(\gamma
_{t}+\gamma _{t})]\sinh (\Gamma _{t,\tau })+C_{pf}^{(r)}|_{\breve{y}}.$ In
these expressions, $\Gamma _{t,\tau }=\gamma _{t}+\gamma _{\tau }-\gamma
_{t+\tau }$ and $\Phi _{t,\tau }=\phi _{t}+\phi _{\tau }-\phi _{t+\tau }$
where $\gamma _{t}\equiv 4\sum\nolimits_{k}(|g_{k}|^{2}/\omega
_{k}^{2})[1-\cos (\omega _{k}t)],$ and $\phi _{t}\equiv
4\sum\nolimits_{k}(|g_{k}|^{2}/\omega _{k}^{2})\sin (\omega _{k}t).$

Assuming the spectral density $J(\omega )=\lambda \omega \exp (-\omega
/\omega _{c}),$ where $\omega _{c}$ is a cutoff frequency \cite{breuerbook},
it follows $\gamma _{t}=(1/2)\ln [1+(\omega _{c}t)^{2}]$ and $\phi
_{t}=\arctan [\omega _{c}t]$ ($\lambda =1$). In Fig.~1(c) we plot the CPF
correlation of both schemes. Even when the unperturbed system dynamics can
be written as a (continuous) statistical superposition of unitary dynamics 
\cite{nori}, our approach detects the presence of BIFs, $C_{pf}^{(r)}|_{%
\breve{y}}\neq 0.$ In fact, $C_{pf}^{(r)}|_{\breve{y}}=0$ only occurs for
very specific measurement directions.

\section{Conclusions}

Memory effects in open quantum systems may underlay or not on a
bidirectional system-environment physical exchange of information. We
introduced an operational scheme that allow to distinguishing between both
situations, solving a long standing problem in the theory of non-Markovian
open quantum systems. The method is based on a probabilistic relation that
relates the developing of BIFs with the modification of the environmental
dynamical behavior. We showed that BIFs can be detected with a minimal
number of three system measurement processes added to an intermediate system
update operation.

A generalized CPF correlation, defined between the first and last
measurement outcomes, witnesses memory effects. Depending on the system
state update scheme, deterministic vs. random, it witnesses memory effects
independently of its underlying origin or restricted to the presence of BIFs
respectively. Consistently, for environments modeled by classical noise
fluctuations or when the environment dynamics (incoherent or quantum) is not
affected during the system evolution, not any BIFs is detected. The presence
of BIFs for decay and dephasing dynamics modeled through unitary
system-environment interactions also support the consistence of the
developed approach.

Given the operational character of the proposed scheme, it can be
implemented, for example, in quantum optical arrangements \cite%
{budiniChina,budiniBrasil}, providing in general a valuable experimental
tool for studying the underlying origin of quantum memory effects.
Generalizations for an arbitrary number of measurement processes can also be
worked out in a similar way. The proposed theoretical ground may also shed
light on the possibility of classifying memory effects in classical and
quantum ones \cite{costa}, and may also provide an explicit test for
different (causal) structures arising in quantum causal modelling~\cite%
{causal}.

\section*{Acknowledgments}

This paper was supported by Consejo Nacional de Investigaciones Cient\'{\i}%
ficas y T\'{e}cnicas (CONICET), Argentina.

\appendix

\section{Joint probabilities}

The system is subjected to three measurement processes performed at times $%
0\rightarrow t\rightarrow t+\tau .$ The corresponding measurement operators
are denoted as $\{\Omega _{x}\},$ $\{\Omega _{y}\},$ and $\{\Omega _{z}\}.$
The intermediate $y$-measurement is taken as a projective one, $\Omega
_{y}=|y\rangle \langle y|.$ The corresponding post-measurement system state
is $\rho _{y}=|y\rangle \langle y|.$ After this step, the state
transformation $\rho _{y}\rightarrow \rho _{\breve{y}}$ is externally
applied. Each of the possible states $\{\rho _{\breve{y}}\}$ is chosen with
conditional probability $\wp (\breve{y}|y,x),$ which only depends on the
previous particular measurement outcomes $x$ and $y.$

The relevant joint probability $P(z,\breve{y},x)$ for the present proposal
can be obtained as%
\begin{equation}
P(z,\breve{y},x)=\sum_{y}P(z,\breve{y},y,x).
\end{equation}%
The joint probability for the four events $P(z,\breve{y},y,x)$ follows from
standard quantum measurement theory after knowing the open system dynamics.
The CPF probability $P(z,x|\breve{y}),$ which determine the CPF correlation
[Eq.~(\ref{GCPF})] \cite{budiniCPF}, can straightforwardly be obtained as%
\begin{equation}
P(z,x|\breve{y})=P(z,\breve{y},x)/P(\breve{y}),  \label{CPFProbability}
\end{equation}%
where $P(\breve{y})=\sum_{z,x}P(z,\breve{y},x)=\sum_{z,y,x}P(z,\breve{y}%
,y,x).$ In addition, $P(z|\breve{y})=\sum_{x}P(z,x|\breve{y})$ and $P(x|%
\breve{y})=\sum_{z}P(z,x|\breve{y}).$

\subsection{Classical noise environmental fluctuations}

For classical noisy environments the outcomes probabilities are obtained for
each realization, while an ensemble average is performed at the end of the
calculation.

Let $\rho _{0}$ denotes the initial system state. After performing the first
system measurement, with operators $\{\Omega _{x}\},$ it occurs the
transformation $\rho _{0}\rightarrow \rho _{x},$ where%
\begin{equation}
\rho _{x}=\frac{\Omega _{x}\rho _{0}\Omega _{x}^{\dagger }}{\mathrm{Tr}%
_{s}(E_{x}\rho _{0})}.  \label{EmeX}
\end{equation}%
Here, $E_{x}=\Omega _{x}^{\dagger }\Omega _{x}.$ The probability of each
outcome is%
\begin{equation}
P(x)=\mathrm{Tr}_{s}(E_{x}\rho _{0}).  \label{PxST}
\end{equation}%
During the time interval$\ 0\rightarrow t,$ the system evolves with a
(completely positive) dynamics defined by the stochastic propagator $%
\mathcal{E}_{t,0}^{st}.$ After the second $y$-measurement, with operators $%
\{\Omega _{y}\},$ it follows the transformation $\mathcal{E}_{t,0}^{st}[\rho
_{x}]\rightarrow \rho _{y},$ where%
\begin{equation}
\rho _{y}=\frac{\Omega _{y}\mathcal{E}_{t,0}^{st}[\rho _{x}]\Omega
_{y}^{\dagger }}{\mathrm{Tr}_{s}(E_{y}\mathcal{E}_{t,0}^{st}[\rho _{x}])}%
=|y\rangle \langle y|,
\end{equation}%
and $E_{y}=\Omega _{y}^{\dagger }\Omega _{y}.$ Here, we used that the $y$%
-measurement is a projective one, $\Omega _{y}=|y\rangle \langle y|.$ The
conditional probability $P^{st}(y|x)$ of outcome $y$ given that the previous
one was $x$ is%
\begin{equation}
P^{st}(y|x)=\mathrm{Tr}_{s}(E_{y}\mathcal{E}_{t,0}^{st}[\rho _{x}]).
\label{P(y|x)ST}
\end{equation}%
At this stage, independently of the outcome $y,$ the system state is updated
as $\rho _{y}\rightarrow \rho _{\breve{y}}.$ The states $\{\rho _{\breve{y}%
}\}$ are chosen with conditional probability $\wp (\breve{y}|y,x),$ which
does not depend on the particular noise realization.

In the final steps $(t\rightarrow t+\tau ),$ the system evolves with the
propagator $\mathcal{E}_{t+\tau ,t}^{st}$ and the last $z$-measurement, with
operators $\{\Omega _{z}\},$ is performed ($\tau $ is the time interval
between the measurements). Thus, $\mathcal{E}_{t+\tau ,t}^{st}[\rho _{\breve{%
y}}]\rightarrow \rho _{z}^{st},$ where%
\begin{equation}
\rho _{z}^{st}=\frac{\Omega _{z}\mathcal{E}_{t+\tau ,t}^{st}[\rho _{\breve{y}%
}]\Omega _{z}^{\dagger }}{\mathrm{Tr}_{s}(E_{z}\mathcal{E}_{t+\tau
,t}^{st}[\rho _{\breve{y}}])},
\end{equation}%
with $E_{z}=\Omega _{z}^{\dagger }\Omega _{z}.$ The conditional probability
of outcome $z$ given that the previous ones were $x$ and $y,$ and given that
the state $\rho _{\breve{y}}$ was imposed, is%
\begin{equation}
P^{st}(z|\breve{y},y,x)=\mathrm{Tr}_{s}(E_{z}\mathcal{E}_{t+\tau
,t}^{st}[\rho _{\breve{y}}]).  \label{P(z|yyx)ST}
\end{equation}%
For each noise realization, this object does not depend on outcomes $y$ and $%
x.$

The joint probability of the four events $P(z,\breve{y},y,x)$ can be
obtained as an average over an ensemble of realizations. Denoting the
average operation with the overbar symbol, Bayes rule leads to%
\begin{equation}
P(z,\breve{y},y,x)=\overline{P^{st}(z|\breve{y},y,x)\wp (\breve{y}%
|y,x)P^{st}(y|x)P(x)}.
\end{equation}%
From Eqs. (\ref{PxST}), (\ref{P(y|x)ST}), and (\ref{P(z|yyx)ST}), we get%
\begin{equation}
\frac{P(z,\breve{y},y,x)}{\wp (\breve{y}|y,x)}=\overline{\mathrm{Tr}%
_{s}(E_{z}\mathcal{E}_{t+\tau ,t}^{st}[\rho _{\breve{y}}])\mathrm{Tr}%
_{s}(E_{y}\mathcal{E}_{t,0}^{st}[\tilde{\rho}_{x}])},  \label{P4_Noise}
\end{equation}%
where $\tilde{\rho}_{x}\equiv \Omega _{x}\rho _{0}\Omega _{x}^{\dagger },$
which recovers Eq.~(\ref{PJoint4StochHamiltonian}).

\subsection{Completely positive system-environment dynamics}

Let $\rho _{0}^{se}=\rho _{0}\otimes \sigma _{0}$ denotes the bipartite
state at the initial time. After performing the first system measurement,
with operators $\{\Omega _{x}\},$ it occurs the transformation $\rho
_{0}^{se}\rightarrow \rho _{x}^{se},$ where the post-measurement state is%
\begin{equation}
\rho _{x}^{se}=\frac{\Omega _{x}\rho _{0}^{se}\Omega _{x}^{\dagger }}{%
\mathrm{Tr}_{se}(E_{x}\rho _{0}^{se})},
\end{equation}%
with $E_{x}=\Omega _{x}^{\dagger }\Omega _{x}.$ The probability of each
outcome is%
\begin{equation}
P(x)=\mathrm{Tr}_{s}(E_{x}\rho _{0}).  \label{P(x)}
\end{equation}%
During the time interval$\ 0\rightarrow t,$ the bipartite arrangement
evolves with a completely positive dynamics defined by the propagator $%
\mathcal{E}_{t,0}.$ After the second $y$-measurement, it follows the
transformation $\mathcal{E}_{t,0}[\rho _{x}^{se}]\rightarrow \rho _{y}^{se},$
where%
\begin{equation}
\rho _{y}^{se}=\frac{\Omega _{y}\mathcal{E}_{t,0}[\rho _{x}^{se}]\Omega
_{y}^{\dagger }}{\mathrm{Tr}_{se}(E_{y}\mathcal{E}_{t}[\rho _{x}^{se}])}%
=\rho _{y}\otimes \sigma _{e}^{yx}.  \label{RhoYBipartito}
\end{equation}%
Here, $E_{y}=\Omega _{y}^{\dagger }\Omega _{y}.$ In the last equality we
used that the second measurement is a projective one, $\Omega _{y}=|y\rangle
\langle y|$ and $\rho _{y}=|y\rangle \langle y|.$ The environment state is%
\begin{equation}
\sigma _{e}^{yx}=\frac{\mathrm{Tr}_{s}(E_{y}\mathcal{E}_{t,0}[\rho
_{x}^{se}])}{\mathrm{Tr}_{se}(E_{y}\mathcal{E}_{t,0}[\rho _{x}^{se}])}.
\label{BathStateYX}
\end{equation}%
The conditional probability $P(y|x)$ of outcome $y$ given that the previous
one was $x$ is%
\begin{equation}
P(y|x)=\mathrm{Tr}_{se}(E_{y}\mathcal{E}_{t,0}[\rho _{x}^{se}]).
\label{P(y|x)}
\end{equation}

At this stage, independently of the outcome $y,$ the system is initialized
in an independently chosen state $\rho _{\breve{y}},$ with conditional
probability $\wp (\breve{y}|y,x).$ Thus, the bipartite state [Eq.~(\ref%
{RhoYBipartito})] becomes%
\begin{equation}
\rho _{y}^{se}\rightarrow \rho _{\breve{y}}^{se}=\rho _{\breve{y}}\otimes
\sigma _{e}^{yx}.
\end{equation}%
In the final steps $(t\rightarrow t+\tau ),$ the bipartite system
arrangement evolves with the propagator $\mathcal{E}_{t+\tau ,t},$ and the
last $z$-measurement is performed. Hence, $\mathcal{E}_{t+\tau ,t}[\rho _{%
\breve{y}}\otimes \sigma _{e}^{yx}]\rightarrow \rho _{z}^{se},$ where%
\begin{equation}
\rho _{z}^{se}=\frac{\Omega _{z}\mathcal{E}_{t+\tau ,t}[\rho _{\breve{y}%
}\otimes \sigma _{e}^{yx}]\Omega _{z}^{\dagger }}{\mathrm{Tr}_{se}(E_{z}%
\mathcal{E}_{t+\tau ,t}[\rho _{\breve{y}}\otimes \sigma _{e}^{yx}])},
\label{RhoZAp}
\end{equation}%
with $E_{z}=\Omega _{z}^{\dagger }\Omega _{z}.$ The conditional probability
of outcome $z$ given that the previous ones were $x$ and $y,$ and given that
the state $\rho _{\breve{y}}$ was imposed, is%
\begin{equation}
P(z|\breve{y},y,x)=\mathrm{Tr}_{se}(E_{z}\mathcal{E}_{t+\tau ,t}[\rho _{%
\breve{y}}\otimes \sigma _{e}^{yx}]).  \label{PZconditionalYTilde}
\end{equation}

From Bayes rule, the joint probability $P(z,\breve{y},y,x)$\ of the four
events can be written as%
\begin{equation}
P(z,\breve{y},y,x)=P(z|\breve{y},y,x)\wp (\breve{y}|y,x)P(y|x)P(x).
\end{equation}%
From Eqs. (\ref{P(x)}), (\ref{P(y|x)}), and (\ref{PZconditionalYTilde}),\ it
follows%
\begin{eqnarray}
P(z,\breve{y},y,x) &=&\mathrm{Tr}_{se}(E_{z}\mathcal{E}_{t+\tau ,t}[\rho _{%
\breve{y}}\otimes \sigma _{e}^{yx}])  \notag \\
&&\wp (\breve{y}|y,x)\mathrm{Tr}_{se}(E_{y}\mathcal{E}_{t,0}[\tilde{\rho}%
_{x}^{se}]),  \label{preFormula}
\end{eqnarray}%
where $\tilde{\rho}_{x}^{se}\equiv \Omega _{x}\rho _{0}^{se}\Omega
_{x}^{\dagger }.$ Using Eq. (\ref{BathStateYX}) for $\sigma _{e}^{yx},$
finally we get%
\begin{equation}
\frac{P(z,\breve{y},y,x)}{\wp (\breve{y}|y,x)}=\mathrm{Tr}_{se}(E_{z}%
\mathcal{E}_{t+\tau ,t}[\rho _{\breve{y}}\otimes \mathrm{Tr}_{s}(E_{y}%
\mathcal{E}_{t,0}[\tilde{\rho}_{x}^{se}])]),  \label{P4QuantumBipartite}
\end{equation}%
which recovers Eq.~(\ref{Prob4CP}).

\subsection{Unconditional environment average}

The calculus of $P(z,\breve{y},y,x)$ in the previous section relies on the
association $\langle \cdots \rangle _{\mathbf{e}}\leftrightarrow \mathrm{Tr}%
_{e}(\cdots ).$ This unconditional environment average emerges when the the
successive (non-selective) measurement of the environment do not modify its
state at each stage. While this result follows straightforwardly from
quantum measurement theory \cite{breuerbook}, here it is explicitly
confirmed.

We consider three measurement processes but now they provide information of
both the system and the environment. The successive outcomes are denoted as $%
x\rightarrow (y\rightarrow \breve{y})\rightarrow z$ and $\mathfrak{X}%
\rightarrow \mathfrak{Y}\rightarrow \mathfrak{Z}$ (Latin and Fraktur
letters) respectively. Introducing the notation $X=(x,\mathfrak{X}),$ $Y=(y,%
\mathfrak{Y}),$ and $Z=(z,\mathfrak{Z}),$ the measurement operators are
denoted as $\{\Omega _{X}\},$ $\{\Omega _{Y}\},$ and $\{\Omega _{Z}\},$
where $\Omega _{X}=\Omega _{x}\otimes \Omega _{\mathfrak{X}},$ $\Omega
_{Y}=\Omega _{y}\otimes \Omega _{\mathfrak{Y}}$ and $\Omega _{Z}=\Omega
_{z}\otimes \Omega _{\mathfrak{Z}}.$ As before, the intermediate system
measurement is taken as a projective one, $\Omega _{y}=|y\rangle \langle y|.$

From Bayes rule, the probability of all measurements and preparation events
can be written as%
\begin{equation}
P(Z,\breve{y},Y,X)=P(Z|\breve{y},Y,X)\wp (\breve{y}|y,x)P(Y|X)P(X).
\end{equation}%
By performing the same calculus steps as in the previous section, from Eqs. (%
\ref{preFormula}) straightforwardly we obtain%
\begin{eqnarray}
P(Z,\breve{y},Y,X) &=&\mathrm{Tr}_{se}(E_{Z}\mathcal{E}_{t+\tau ,t}[\rho _{%
\breve{y}}\otimes \sigma _{e}^{YX}])  \notag \\
&&\wp (\breve{y}|y,x)\mathrm{Tr}_{se}(E_{Y}\mathcal{E}_{t,0}[\tilde{\rho}%
_{X}^{se}]),  \label{4Previa}
\end{eqnarray}%
where $E_{J}=\Omega _{J}^{\dagger }\Omega _{J}$ $(J=X,Y,Z),$ and $\tilde{\rho%
}_{X}^{se}=\Omega _{X}\rho _{0}^{se}\Omega _{X}^{\dagger }.$\ Furthermore,%
\begin{equation}
\sigma _{e}^{YX}=\frac{\Omega _{\mathfrak{Y}}\mathrm{Tr}_{s}(E_{y}\mathcal{E}%
_{t,0}[\rho _{X}^{se}])\Omega _{\mathfrak{Y}}^{\dag }}{\mathrm{Tr}_{se}(E_{Y}%
\mathcal{E}_{t,0}[\rho _{X}^{se}])},
\end{equation}%
where $\rho _{X}^{se}=\tilde{\rho}_{X}^{se}/\mathrm{Tr}_{se}(E_{X}\rho
_{0}^{se}).$ Similarly, Eq. (\ref{4Previa}) can be rewritten as%
\begin{equation}
\!\frac{P(\!Z,\!\breve{y},\!Y,\!X\!)}{\wp (\breve{y}|y,x)}\!=\!\mathrm{Tr}%
_{se}(\!E_{Z}\mathcal{E}_{t+\tau ,t}[\rho _{\breve{y}}\otimes \Omega _{%
\mathfrak{Y}}\mathrm{Tr}_{s}(E_{y}\mathcal{E}_{t,0}[\tilde{\rho}%
_{X}^{se}])\Omega _{\mathfrak{Y}}^{\dag }]).  \label{4DobleFinal}
\end{equation}

The probability for the environment outcomes\ follows by marginating the
system outcomes,%
\begin{equation}
P(\mathfrak{Z},\mathfrak{Y},\mathfrak{X})=\sum_{z,\breve{y},y,x}P(Z,\breve{y}%
,Y,X).
\end{equation}%
Similarly, the probability for the system outcomes\ follows by marginating
the outcomes corresponding to the reservoir measurements,%
\begin{equation}
P(z,\breve{y},y,x)=\sum_{\mathfrak{Z},\mathfrak{Y},\mathfrak{X}}P(Z,\breve{y}%
,Y,X).  \label{P4MedisBath}
\end{equation}%
This result for $P(z,\breve{y},y,x)$ relies on explicit environment
measurements. In contrast, the results of the previous section were derived
assuming that the environment is not observed at all. Nevertheless, both
kind of results can be put in one-to-one correspondence. In fact, Eqs.~(\ref%
{preFormula}) and (\ref{P4QuantumBipartite}) can be recovered from Eqs. (\ref%
{4Previa}) and (\ref{4DobleFinal}), via the margination (\ref{P4MedisBath}),
under the conditions%
\begin{equation}
\sigma _{0}=\sum_{\mathfrak{X}}\Omega _{\mathfrak{X}}\sigma _{0}\Omega _{%
\mathfrak{X}}^{\dag },\ \ \ \ \ \ \ \sigma _{e}^{yx}=\sum_{\mathfrak{Y}%
}\Omega _{\mathfrak{Y}}\sigma _{e}^{yx}\Omega _{\mathfrak{Y}}^{\dag },
\label{invariancia}
\end{equation}%
where $\sigma _{0}$ is the initial bath state and $\sigma _{e}^{yx}$ is
defined by Eq.~(\ref{BathStateYX}). As expected,\textit{\ these equalities
imply that the bath states at each stage are not modified by the
corresponding reservoir (non-selective) measurement processes.} Thus, the
unconditional environment average of the previous section [Eq.~(\ref%
{P4QuantumBipartite})] relies on this kind of observables, which allow us to
formulate the full approach without performing any explicit reservoir
measurement.

For projective environment measurements, the relations (\ref{invariancia})
implies the commutation relations $[\sigma _{0},\Omega _{\mathfrak{X}}]=0,$ $%
[\sigma _{e}^{yx},\Omega _{\mathfrak{Y}}]=0.$ In classical (incoherent)
reservoirs, where the bath state is diagonal in (a unique) privileged basis,
these conditions define the corresponding \textquotedblleft classical
environment observables.\textquotedblright

\section{Eternal non-Markovianity}

The non-Markovian system density matrix evolution is given by Eq. (\ref%
{MasterEternal}). There exist different underlying dynamic that lead to this
dynamics. The solution map $\rho _{0}\rightarrow \rho _{t}$ can be written
as a mixture of three Markovian maps \cite{megier}%
\begin{equation}
\rho _{t}=\sum_{\alpha =\hat{x},\hat{y},\hat{z}}q_{\alpha }\mathcal{E}%
_{t,0}^{(\alpha )}[\rho _{0}],
\end{equation}%
with positive and normalized statistical weights $\{q_{\alpha }\},$ $%
\sum_{\alpha =\hat{x},\hat{y},\hat{z}}q_{\alpha }=1.$ The Markovian
propagators are%
\begin{equation}
\mathcal{E}_{t,t_{0}}^{(\alpha )}[\rho _{0}]=h_{t-t_{0}}^{(+)}\rho
_{0}+h_{t-t_{0}}^{(-)}\sigma _{\alpha }\rho _{0}\sigma _{\alpha },
\label{PropaAlfa}
\end{equation}%
with scalar functions $h_{t}^{(\pm )}\equiv (1\pm e^{-2\gamma t})/2.$ Each
propagator $\mathcal{E}_{t,t_{0}}^{(\alpha )}=\exp [\gamma (t-t_{0})\mathbb{L%
}_{\alpha }]$ corresponds to the solution of the Markovian Lindblad evolution%
\begin{equation}
\frac{d}{dt}=\gamma \mathbb{L}_{\alpha }[\rho _{t}]=\gamma (\sigma _{\alpha
}\rho _{t}\sigma _{\alpha }-\rho _{t}).
\end{equation}%
The evolution (\ref{MasterEternal}) emerges with $q_{\hat{x}}=q_{\hat{y}%
}=1/2 $ and $q_{\hat{z}}=0$ \cite{megier}.

The probability $P(z,\breve{y},y,x)$ can be straightforwardly be obtained
from Eq. (\ref{P4_Noise}) under the replacement $\overline{(\cdots )}%
\rightarrow \sum_{\alpha }q_{\alpha }\cdots .$ We get%
\begin{equation}
\frac{P(z,\breve{y},y,x)}{\wp (\breve{y}|y,x)}=\!\sum_{\alpha =\hat{x},\hat{y%
},\hat{z}}\!\!\!q_{\alpha }\mathrm{Tr}_{s}(E_{z}\mathcal{E}_{t+\tau
,t}^{(\alpha )}[\rho _{\breve{y}}])\mathrm{Tr}_{s}(E_{y}\mathcal{E}%
_{t,0}^{(\alpha )}[\tilde{\rho}_{x}]),  \label{P4eternal}
\end{equation}%
where $E_{i}\equiv \Omega _{i}^{\dagger }\Omega _{i}$ $(i=x,y,z)$ and $%
\tilde{\rho}_{x}\equiv \Omega _{x}\rho _{0}\Omega _{x}^{\dagger }$ is the
(unnormalized) system state after the first $x$-measurement. $\mathrm{Tr}%
_{s}(\cdots )$ denotes a trace operation in the system Hilbert space. $\rho
_{\breve{y}}$ is the (updated) system state after the second $y$-measurement.

In the deterministic scheme $[\wp (\breve{y}|y,x)=\delta _{\breve{y},y}],$
using that $P(z,\breve{y},x)=\sum_{y}P(z,\breve{y},y,x),$ Eq. (\ref%
{P4eternal}) leads to%
\begin{equation}
P(z,\breve{y},x)\overset{d}{=}\sum_{\alpha =\hat{x},\hat{y},\hat{z}%
}q_{\alpha }\mathrm{Tr}_{s}(E_{z}\mathcal{E}_{t+\tau ,t}^{(\alpha )}[\rho _{%
\breve{y}}])\mathrm{Tr}_{s}(E_{\breve{y}}\mathcal{E}_{t,0}^{(\alpha )}[%
\tilde{\rho}_{x}]).  \label{PZXEternal}
\end{equation}%
In general, this joint probability does not fulfill a Markov condition.
Thus, $C_{pf}^{(d)}|_{\breve{y}}\neq 0$ detects memory effects. On the other
hand, in the random scheme $[\wp (\breve{y}|y,x)=\wp (\breve{y}|x)]$ it
follows%
\begin{equation}
P(z,\breve{y},x)\overset{r}{=}\sum_{\alpha =\hat{x},\hat{y},\hat{z}%
}q_{\alpha }\mathrm{Tr}_{s}(E_{z}\mathcal{E}_{t+\tau ,t}^{(\alpha )}[\rho _{%
\breve{y}}])\wp (\breve{y}|x)\mathrm{Tr}_{s}(\tilde{\rho}_{x}),
\label{ProbRandomEternal}
\end{equation}%
which recovers a Markovian structure, $P(z,\breve{y},x)=P(z|\breve{y})\wp (%
\breve{y}|x)P(x),$ with $P(z|\breve{y})=\sum_{\alpha =\hat{x},\hat{y},\hat{z}%
}q_{\alpha }\mathrm{Tr}_{s}(E_{z}\mathcal{E}_{t+\tau ,t}^{(\alpha )}[\rho _{%
\breve{y}}])$ and $P(x)=\mathrm{Tr}_{s}(\tilde{\rho}_{x}).$ Thus,
independently of the chosen system measurement observables it follows $%
C_{pf}^{(r)}|_{\breve{y}}=0,$ indicating consistently the absence of any BIF.

\subsection*{$\hat{x}$-$\hat{n}$-$\hat{x}$ measurements}

We consider the case in which the three measurements are projective ones.
The first and third ones are performed in $\hat{x}$-direction of the Bloch
sphere. The intermediate one is performed in a direction $\hat{n}=\{\sin
(\theta ),0,\cos (\theta )\},$ which lies in the $\hat{x}$-$\hat{z}$ plane
of the Bloch sphere. Thus, the measurement operators are $\Omega _{x=\pm }=|%
\hat{x}_{\pm }\rangle \langle \hat{x}_{\pm }|,$ $\Omega _{y=\pm }=|\hat{n}%
_{\pm }\rangle \langle \hat{n}_{\pm }|,$ and $\Omega _{z=\pm }=|\hat{x}_{\pm
}\rangle \langle \hat{x}_{\pm }|.$ Consistently with the chosen directions,\
we have $|\hat{x}_{\pm }\rangle =(|+\rangle \pm |-\rangle )/\sqrt{2},$
jointly with $|\hat{n}_{+}\rangle =\cos (\theta /2)|+\rangle +\sin (\theta
/2)|-\rangle ,$ and $|\hat{n}_{-}\rangle =\sin (\theta /2)|+\rangle -\cos
(\theta /2)|-\rangle .$

For an explicit calculation of the previous probabilities we need to
calculate $P_{\alpha }(\hat{n}|\hat{x})\equiv \mathrm{Tr}_{s}(E_{\hat{n}}%
\mathcal{E}_{t_{f},t_{i}}^{(\alpha )}[\rho _{\hat{x}}])$ and $P_{\alpha }(%
\hat{x}|\hat{n})\equiv \mathrm{Tr}_{s}(E_{\hat{x}}\mathcal{E}%
_{t_{f},t_{i}}^{(\alpha )}[\rho _{\hat{n}}]),$ where $E_{\hat{n}}=|\hat{n}%
\rangle \langle \hat{n}|$ and $\rho _{\hat{x}}=|\hat{x}\rangle \langle \hat{x%
}|.$ From Eq. (\ref{PropaAlfa}) and the definition of the measurement
operators, we get%
\begin{equation}
P_{\alpha }(\hat{n}|\hat{x})=P_{\alpha }(\hat{x}|\hat{n}%
)=h_{t_{f}-t_{i}}^{(+)}|\langle \hat{n}|\hat{x}\rangle
|^{2}+h_{t_{f}-t_{i}}^{(-)}|\langle \hat{n}|\sigma _{\alpha }|\hat{x}\rangle
|^{2}.  \label{Palfa}
\end{equation}%
Using this result, after a straightforward calculation, from Eqs.~(\ref%
{PZXEternal}) and (\ref{Palfa}) we get%
\begin{eqnarray}
&&P(z,\breve{y},x)\overset{d}{=}\frac{1}{4}[1+\breve{y}x\sin (\theta )c(t)+z%
\breve{y}\sin (\theta )c(\tau )  \notag \\
&&\ \ \ \ \ \ \ \ \ \ \ \ \ \ \ \ \ \ +zx\sin ^{2}(\theta )c(t+\tau )]P(x),
\label{P3DeterministaEternal}
\end{eqnarray}%
where $P(x)=\mathrm{Tr}_{s}(E_{x}\rho _{0}),$ and%
\begin{equation}
c(t)\equiv q_{\hat{x}}+(q_{\hat{y}}+q_{\hat{z}})\exp [-2\gamma t].
\end{equation}%
In the random scheme, from Eq. (\ref{ProbRandomEternal}) we obtain%
\begin{equation}
P(z,\breve{y},x)\overset{r}{=}\frac{1}{2}[1+z\breve{y}\sin (\theta )c(\tau
)]\wp (\breve{y}|x)P(x),  \label{P3RandomEternal}
\end{equation}%
where we considered the updated states $\rho _{\breve{y}=\pm 1}=|\hat{n}%
_{\pm }\rangle \langle \hat{n}_{\pm }|.$

The generalized CPF correlation is given by Eq. (\ref{GCPF}), $%
C_{pf}^{(d/r)}|_{\breve{y}}=\sum_{zx}O_{z}O_{x}[P(z,x|\breve{y})-P(z|\breve{y%
})P(x|\breve{y})],$ where $P(z,x|\breve{y})$ follows from Eq. (\ref%
{CPFProbability}). Furthermore, $O_{z}=z=\pm 1$ and $O_{x}=x=\pm 1.$ From
Eq. (\ref{P3DeterministaEternal}), the CPF correlation in the deterministic
scheme reads%
\begin{equation}
C_{pf}^{(d)}|_{\breve{y}}\underset{\hat{x}\hat{n}\hat{x}}{=}\sin ^{2}(\theta
)\frac{[1-\langle x\rangle ^{2}]}{4[P(\breve{y})]^{2}}[c(t+\tau )-c(t)c(\tau
)],
\end{equation}%
where $P(\breve{y})=(1/2)[1+\breve{y}\langle x\rangle \sin (\theta )c(t)]$
and $\langle x\rangle \equiv \sum_{x=\pm 1}xP(x).$ When $\rho _{0}=|\pm
\rangle \langle \pm |$ it follows $P(x)=1/2$ and consequently $\langle
x\rangle =0.$ This case recovers Eq. (\ref{CPFDetEternal}).

In the random scheme, from Eq. (\ref{P3RandomEternal}) consistently it
follows%
\begin{equation}
C_{pf}^{(r)}|_{\breve{y}}=0.
\end{equation}%
This equality is valid independently of the chosen measurement processes and
updated system states [see Eq.~(\ref{ProbRandomEternal})].


\begin{thebibliography}{99}
\bibitem{repraration} Formally, the update $\rho _{y}\rightarrow \rho _{%
\breve{y}}$ is equivalent to discard the system state and feeds forward an
independent system state. Named as \textquotedblleft
repreparation,\textquotedblright\ this operation was\ introduced in Ref.~%
\cite{pollock} for studying \textquotedblleft quantum Markov
order.\textquotedblright\ The formulation with unitary transformations $\{U(%
\breve{y}|y)\}$ defines a feasible experimental implementation.
\bibitem{invariance} Under the condition $[H_{e},H_{I}]=0,$ the invariance
property (\ref{Nonsignalin}) is fulfilled only when the initial bath state
satisfies $[\sigma _{0},H_{e}]=0.$
\end{thebibliography}
\end{document}